# Dynamic Analysis of Confined Ionic Liquids and Ionic Magnetic Fluids: Highlighting Long Range Elastic Interactions

Eni Kume, Nicolas Martin, Peter Dunne, Patrick Baroni, Laurence Noirez*

Laboratoire Léon Brillouin (CEA-CNRS), Univ. Paris-Saclay, 91191 Gif-sur-Yvette Cedex, France

Abstract: By reinforcing the interaction energy of the liquid with respect to the surface using total wetting boundary conditions, the response of liquids to mechanical shear stress is stronger and exhibits at sub-millimeter scale elastic properties. This study extends here to liquids with strong electrostatic interactions such as ionic liquids and paramagnetic liquids. We show that it is also possible to identify and measure a non-zero low-frequency shear elasticity at sub-millimeter scale. The observation of mesoscopic elastic properties in liquids generally considered as viscous away from any phase transition and the absence of extended structuring confirm the relevance of considering elastic interactions as contributing to collective effects under external electric or magnetic field.

# Introduction:

The microfluidic scale is certainly the scale at which the limits of the hydrodynamic conditions are the more noticeable. Introducing non-extensive parameters like the influence of interfacial interactions, the scale or the recently identified mesoscopic shear elasticity of liquids is meaningful. Shear elasticity is an indicator of liquid cohesion which is directly related to the strength of intermolecular interactions. These factors might be even more exacerbated when it comes to describe the properties of charged liquids like ionic liquids which is the scope of the present study.

While solids, liquids and gases possess elastic properties since their density changes under hydrostatic pressure, liquids are not supposed to support shear stresses except at frequency reaching the Maxwell relaxation time $\tau = \eta/G$ where $\eta$ is the fluid viscosity and G the shear modulus [1]. This time scale and modulus should be found in the GHz frequency domain for small molecules corresponding to the inverse of molecular time and moduli on the order of GPa [2]. However, recently a weak but non-negligible shear elasticity has been also highlighted at low frequency and sub-millimeter scale in various ordinary liquids indicating a generic liquid property [3-7]. Its identification in liquids is at variance with conventional hydrodynamics that state that the flow results from viscous liquid properties. It tells that liquid molecules are not dynamically free but long range correlated; i.e. intermolecular interactions are not negligible but provide an elastic character that is typically measurable at sub-millimeter scale.

Examining how a liquid responds to an external field, magnetic or as here mechanical, is an unavoidable way to access a true mechanical characterization of a material. The quality of the characterization depends on the method employed. Recently improved dynamic mechanical methods have been elaborated that prove that liquids propagate shear stress [4-7]. This method is here applied to ionic liquids characterized by strong electrostatic interactions due charge carriers. For this, the fluid is submitted to an oscillatory shear strain of small finite amplitude in accordance with the conventional method of viscoelastic measurements but the transmission of the shear stress is applied using total wetting fluid/surface boundary conditions; i.e. a strong surface energy attracts liquid molecules to the solid surface reducing the occurrence of an interfacial slip [8].

We will show that the small scale elastic property is also valid when electrostatic interactions are strong as in the case of ionic liquids (molten salt), ionic solutions or paramagnetic liquids (solubilized salt) opening the way for another possible understanding of giant collective properties. We will also carry out structural studies by neutron scattering to determine if collective effects are or not, correlated with the formation of clusters.

Ionic liquids have recently gained a lot of interest because of the important progresses in synthesis that proposes new molecules exhibiting liquid properties down to room temperatures. They are generally characterized by a very low surface tension, non-flammability, thermal stability and excellent solvating properties. The understanding of these novel ionic liquids and of their physical properties is a very active investigation field. One of the current debated questions is how to classify these ionic liquids? Conventional liquids [9] or rather glassy-like or liquid crystal materials [10]? In viscous ionic liquids, deviations from a Maxwell behaviour are frequently observed at low frequencies [11] indicating a topological network. Similarly light scattering investigations [12] indicate long range structures reminding the clusters observed by E.W. Fischer [13]. Neutron diffraction [14] and dynamic measurements [15] seem to indicate an undefined frontier between liquid and solid states in term of dynamic and structural correlations similar while having a different (Coulombian) origin for Ionic liquids and liquid crystals.

In this paper, we focus on the liquid phase (away from any phase transition). The chosen ionic liquid is the 1-Ethyl-3-methylimidazolium bis(trifluoromethylsulfonyl)imide ([emim][Tf2N]) which is a molten salt at room temperature. Simulations indicate that the liquid phase of Imidazolium-Based ionic liquids are mainly characterized by cation-anion interactions with a first peak of the structure factor at 4.9-5.3 Å while long range ordering are not expected to extend over 20 Å [16]. In these systems, several structural studies have indicated that only Imidazolium-based ionic liquids with long alkyl chains exhibit the presence of structural peaks. These peaks have been interpreted as due to a smectic-like layering of the alkyl chains [17], Triolo et al. [18] reported observation of a structural peak within the range 0.2 - 0.5 Å$^{-1}$ shifting to lower values of scattering vector with increasing chain length n ≥ 5.8, 9, 13. The studied ionic liquid ([emim][Tf2N]) is made of small molecule complex (Ethyl-3-methylimidazolium) having strong opposite charges. The short alkyl chain and the ion symmetry do not allow any particular intermolecular structuration. Thus, a discussion about the existence of a prepeak is here irrelevant. From a structural point of view, [emim][Tf2N] is a simple liquid.

Ionic interactions can also provide magnetic properties. Paramagnetic properties have been found in liquid metals (at high temperature), molten salts or even aqueous solutions [19]. A paramagnetic ionic liquid contains an inorganic cation that can be a rare earth and exhibits a positive magnetic susceptibility unlike the wide majority of fluids which is diamagnetic. These liquids, called also magnetic ionic liquids, are typically applied for medical imaging as contrast liquid for diagnostic or therapeutic objectives. Paramagnetic liquids have the faculty to exhibit a strong attraction to a magnetic field gradient that defies capillary forces and gravity (insert of Fig.2). The paramagnetic cation (Gd, Tb, Y, etc) does not dissociate from the solution but the whole solution is drained by the magnetic field gradient, demonstrating that a strong coupling between molecules of water and paramagnetic atoms exists. It is calculated that the entropic energy $T\Delta S$ = -1.8 (k.J mol-1) is approximately four orders of magnitude greater than the magnetic energy (1T) of a molar solution [20, 21]. Water molecules being diamagnetic, a high degree of connectivity between liquid molecules and paramagnetic cation is required, which is incompatible in terms of single molecular relaxation times (and uncorrelated density fluctuations). This effect questions about the real nature of this paramagnetic solutions.

# Methods:

The analysis of the dynamic response to a mechanical shear stress provides information on the elastic or the viscous character of materials, and allows quantitative measurements of the viscoelastic character versus frequency and shear strain. The shear stress is generated by placing the sample between two coaxial disks one fixed, the other one coupled to a motor imposing a rotating sinusoidal motion of variable frequency ($\omega$) to the disk. The amplitude of the shear strain γo of amplitude is also variable (strain imposed mode). The second disk is immobile and coupled to a sensor. It measures the stress transmitted by the sample via the torque (σ) transmitted by contact to the disk. Oscillatory motion and torque measurement are provided by a conventional rheometer (Ares II – TA-Instruments). Simultaneously, a 7-digit voltmeter (Keitley; rate: 300 data/s) measures the voltage of the motor imposing the oscillation (input wave associated to the strain amplitude), while another 7-digit voltmeter measures the voltage associated to the sensor (output wave associated to the torque). This setup permits the simultaneous measurement of the shear strain and to the shear stress signals, and of the dynamic profile using the conventional relationship: $\sigma(\omega)= G_o \cdot \gamma_o \cdot \sin(\omega \cdot t + \Delta\varphi)$ with $\sigma(\omega)$ is the shear stress, $G_o$, the shear modulus, $\gamma_o$ the strain amplitude defined as the ratio of the displacement to the sample gap and $\Delta\varphi$ the phase shift between the input and the output waves. This equation can be also expressed in terms of shear elastic (G') and viscous (G") moduli: $\sigma(\omega)=\gamma_o \cdot (G'(\omega) \cdot \sin(\omega \cdot t) + G''(\omega) \cdot \cos(\omega \cdot t))$, with G' the component in phase with the strain, and G" the out of phase component. It should be stressed that the formalism in terms of G' and G" supposes that the resulting stress wave keeps the shape of the imposed strain wave (sinusoidal-like). The study of the wave shape is thus complementary.

As indicated in the introduction, the liquid was confined between two Alumina fixtures to reach total wetting conditions. Due to surface charges (high surface energy), $\alpha$-Alumina increases the interaction energy at the fluid/surface boundary and therefore reinforces the shear stress transmission following the protocol described in [4-6, 24]. Conventional metallic fixtures made generally of Aluminum or Stainless steel does not fulfill full wetting boundary conditions.

Neutron scattering measurements were carried at the Lab. Léon Brillouin (Orphée reactor) to probe the q-scattering range from $10^{-2}$ Å$^{-1}$ up to 3 Å$^{-1}$. A specific beam line was dedicated adapting the 2D neutron detector Barotron (which offers a high spatial resolution [22]) for coupled magnetic-scattering measurements in the intermediate scattering range (5 - 100 Å, thus accessing the first structural liquid peak at 2 Å$^{-1}$). Several specific sample environments were specifically designed to enable the *in situ* measurement of the impact of the magnetic field or of a gradient of a magnetic field (control of the magnetic field intensity, of the field geometry, adaptation of the setups as a function of the instrument) on ionic liquids and paramagnetic liquids. The small-angle neutron spectrometer PA20 [23] was also equipped with a specific magnetic field design to optimize the structural study of the paramagnetic fluids in the low q-range. The selected wavelength and distance were 5 Å and 2 m respectively. The liquids were placed in 1mm diameter capillary tubes for wide angle neutron scattering and in 1mm thick rectangular Hellma cells adapted for small-angle neutron scattering. Neutron spectra were normalized by an incoherent scatterer (Vanadium for wide angle scattering and Plexiglas for small angle scattering).

The ionic liquids (1-Ethyl-3-methylimidazolium bis(trifluoromethylsulfonyl)imide ([emim][Tf2N]) and Gadolinium(III) nitrate) were purchased from Aldrich company. All the experiments were carried out at room temperature.

## Results:

Fig.1a displays the signals over four periods of oscillation of the input sin wave and of the output shear stress of the ionic liquid : 1-Ethyl-3-methylimidazolium bis(trifluoromethylsulfonyl)imide: [emim][Tf2N] measured at 0.100mm gap thickness and 2% strain amplitude. The study of the wave shape shows that the shear strain and the stress waves are almost superposed indicating a nearly instant response of the fluid; i.e. an elastic response to the shear strain.

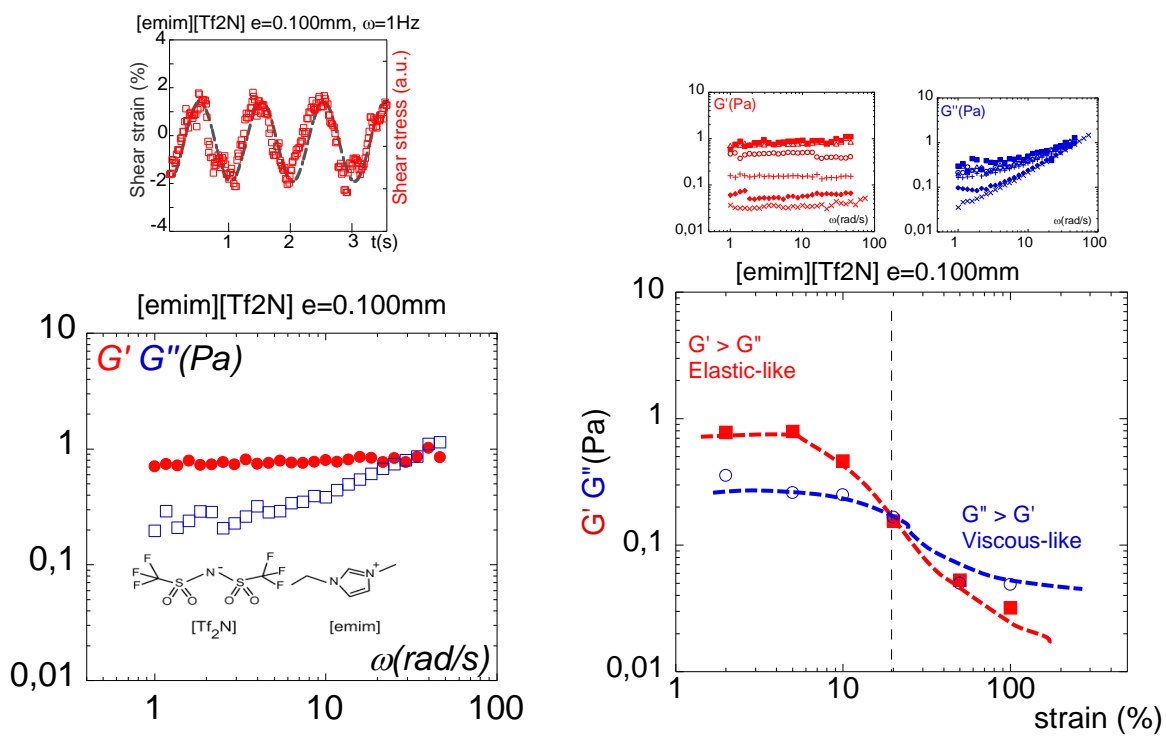

Fig.1a: Top: in-phase (input) shear strain and (output) shear stress waves at 1Hz frequency and 2% shear strain by a 100µm layer of [emim][Tf2N] (1-Ethyl-3-methylimidazolium bis(trifluoromethylsulfonyl)imide).

Bottom: Frequency dependence of the viscoelastic moduli (G', G"). Measurements carried out at 100µm gap thickness, γ = 2% strain amplitude, room temperature and total wetting conditions (Alumina).

Fig.1b: Top: Frequency dependence of the viscoelastic moduli (G', G") displayed by the ionic liquid [emim][Tf2N] at 100µm gap thickness and different strain amplitudes. Bottom: Shear strain induced transition from elastic-like to viscous-like. The shear modulus G' higher than the viscous modulus G' progressively vanishes with increasing strain. Measurements carried out at 2 rad/s, at room temperature using total wetting boundary conditions (alumina). The vertical dashed line indicates the separation line between the elastic and the viscous regimes. The continuous lines are eyeguides.

The evolution of the viscoelastic moduli as a function of the frequency (Fig.1a bottom) confirms the nearly in-phase response of the liquid; the ionic liquid exhibits an elastic behavior recognizable by a shear modulus about one decade higher than a viscous modulus. Therefore, in conditions close to the equilibrium state (the applied shear strain is here 2%), the dynamic mechanical response indicates that the true nature of the ionic liquid is that of a solid characterized by a weak elastic modulus of about the Pascal unit.

Fig.1b shows the dependence of the viscoelastic response as a function of the amplitude of the shear strain. This measurement is also essential since it shows that the elastic response is progressively lost by increasing the shear strain. A transition from elastic-like to viscous-like is observed above 20% shear strain; the ionic liquid exhibits a yield strength, beyond which the measurement indicates a conventional behavior of the viscous type.

The same dynamic mechanical approach is used for paramagnetic liquids. Fig.2 displays the shear stress exhibited by a 0.1M solution of Gadolinium(III) nitrate ($Gd(NO_3)_3$) which displays paramagnetic properties at room temperature. These measurements are carried out at relative large gap thickness (430µm). The dynamic spectrum (Fig.2a) shows that the elastic component is larger than the viscous component, both exhibiting a relative independence with respect to the frequency with in a range from 0.4 rad/s to 10 rad/s. The paramagnetic liquid exhibits in nearly static conditions (1% shear strain) an elastic behavior. This elastic behavior is however very fragile. Fig.2b illustrates the evolution of the elastic-like behavior as a function of the strain amplitude. The shear elasticity is progressively lost over 10% shear strain, highlighting the need to stay close to equilibrium conditions (weak shear strain and low thickness) to probe the elastic-like response. Similarly to the ionic liquid [emim][Tf2N], the large strain plastic behavior makes appear the viscous regime. Since the viscous behavior is obtained at large strain amplitudes, it is the strain induced product of the initial shear elasticity.

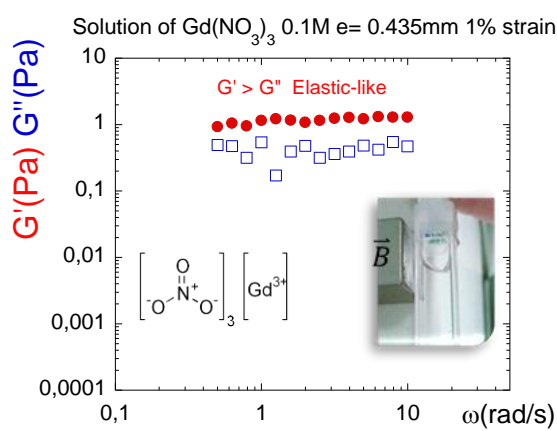
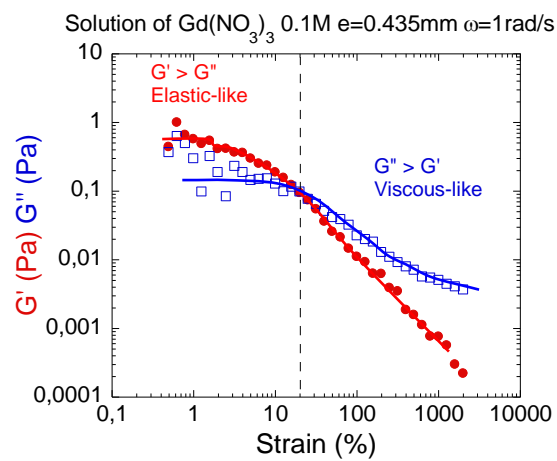

Fig.2a: Low frequency elastic response exhibited by an aqueous solution of 0.1M of $Gd(NO_3)_3$ versus frequency. Room temperature measurements using total wetting conditions (Alumina substrate), sample thickness: 0.430 mm, shear strain: 1%, logarithmic scale. The insert is a picture of a bottle containing the paramagnetic liquid, whose surface is deformed (attracted) by the magnetic field gradient.

Fig.2b: Shear strain induced transition from elastic-like to viscous-like displayed by the 0.1M aqueous solution of $Gd(NO_3)$. The shear modulus G' higher than the viscous modulus G' vanishes with increasing strain making appear the conventional viscous behavior. Measurements carried out at 2 rad/s, at room temperature using total wetting boundary conditions (alumina). The vertical dashed line indicates the separation line between the elastic and the viscous regimes. The continuous lines are eyeguides.

In the last part of this paper, we explore the response of the paramagnetic fluid to a magnetic field viewed from a structural point of view. Paramagnetic solutions exhibit a flow coupling at relatively low magnetic fields (insert of Fig.2a) that challenges conventional thermodynamic considerations. Neutron scattering is a very powerful method able to access bulk structural properties of length scale lying within 5 – 5000 Å due to the exceptional penetration length of the neutron radiation. The strong absorption coefficient of rare earth atoms (Li, Cd, Gd) of these liquids is however a difficulty that excludes the study of high concentrations for bulk samples. To lower the contribution of the incoherent background, the paramagnetic salts are dissolved in deuterated water. Fig.3a illustrates by comparison with heavy water, the intensity scattered by a 0.1M paramagnetic solution of $Tb(NO_3)_3$. The first peak of the structure factor of the paramagnetic solution is very weak compared to the one of the deuterated water, and barely comes out of the background noise even using neutron scattering. Fig.3b compares the scattering of the paramagnetic solution with and without magnetic field (uniform magnetic field of 0.02T). The two scattering curves corresponding to the first peak of the structure factor overlap, making it impossible to identify a magnetic field influence in this Q-range corresponding to the inter-ionic distances. No modification of the diffraction pattern was observed in a magnetic field gradient (tested up to 2T field gradient) or by increasing to concentration to 1M.

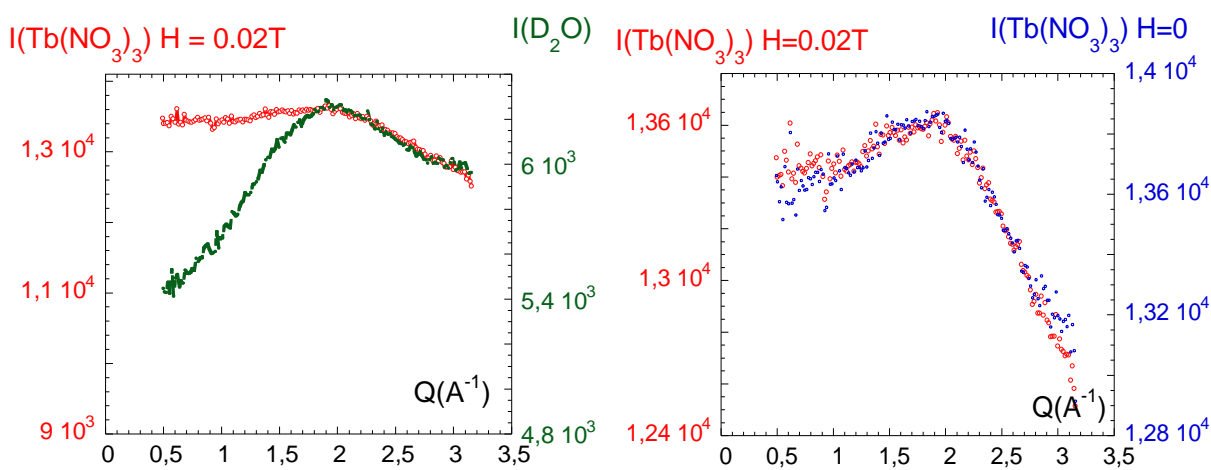

Fig.3a: Wide angle neutron diffraction (Barotron G43, λ = 2.34 Å, d= 0.05m). Heavy paramagnetic atoms are strong radiation absorbers as the comparison between the $D_2O$ scattering and the deuterated solution of $Tb(NO_3)_3$ 0.1M shows. Red points correspond to the paramagnetic solution. Green points correspond to the deuterated water scattering (first peak of the structure factor) and serves as a reference.

Fig.3b: Wide angle neutron diffraction (Barotron G43, λ = 2.34 Å, d= 0.05m). Comparison of the scattering curves corresponding to the first peak of the structure factor of the deuterated solution of $Tb(NO_3)_3$ 0.1M with and without magnetic field. The sample is placed in homogeneous magnetic field oriented perpendicular to the incident beam. Blue points correspond to the scattering without magnetic field while red points correspond to the scattering when the magnetic field was applied (0.02T delivered by an electromagnet).

The small-angle study was carried out on a series of different concentrations of solutions of $Gd(NO_3)_3$ solubilized in deuterated water. The sample was placed in a gradient of a 0.2T magnetic field. Varying the concentration allows to identify a critical solution concentration for which a coupling with the magnetic forces might order the paramagnetic ions in clusters. This test is shown on Fig.4. No structuration could be identified in the in the q range from 0.01 up to 0.1 Å$^{-1}$ within the concentration range up to 0.22 M.

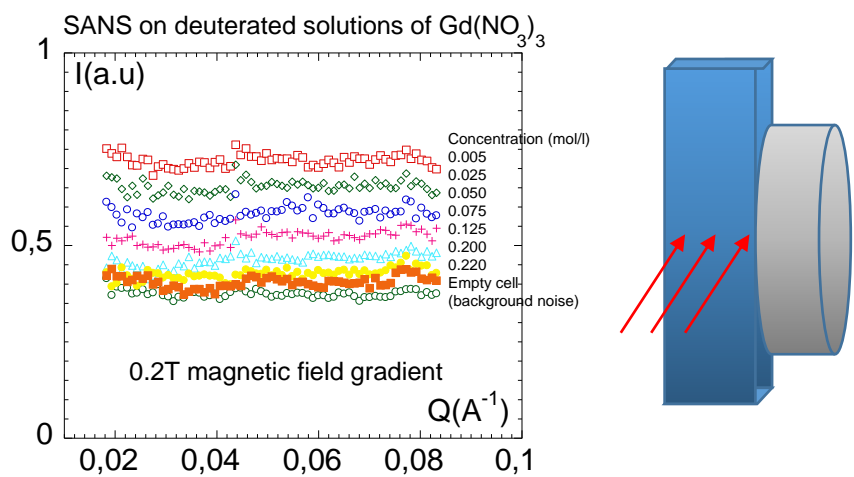

Fig.4: Intensity measured at small angle scattering on a series of different concentrations of solutions of Gd(NO$_3$)$_3$ solubilized in deuterated water. The sample was placed in contact to a pole of 0.2T magnetic field as illustrated in the scheme at right of the figure to submit the paramagnetic liquid to a field gradient. Room temperature measurements.

As consequence, on the basis of a wide Q-range analysis covering scales from 0.01 to 2 Å$^{-1}$ and a series of concentrations, we are able to conclude that the magnetic properties do not contribute significantly to a structuration of Gd(NO$_3$)$_3$ up to 0.22M or Tb(NO$_3$)$_3$ paramagnetic solutions up to concentrations of 1M (higher concentrations are not accessible due to the high absorption coefficients of these rare earths characterized by absorption cross sections of 23 barns and 49700 barns at 1.798 Å for the Terbium and the Gadolinium respectively (NIST data base)). The solutions exhibit typical scattering pattern of simple (ideal) liquids. The magnetic field (or gradient of magnetic field) does not induce any structural change. This result might find the same explanation as for ionic liquids that require long chain to allow a local ordering. Correlatively, magnetically induced collective properties might find some tracks of interpretation related to elastic liquid properties described above (Fig.2).

Conclusions:

We have highlighted that both [emim][Tf2N] and the paramagnetic liquid exhibit a collective response in response to a low frequency mechanical shear strain; An elastic-like response is observed in nearly static strain conditions at low gap thicknesses (0.100mm and 0.435mm). This result corroborates the identification of low frequency elastic behaviour reported on another mesoscopic ionic liquid [24]. The existence of elastic correlations in a liquid phase away from any phase transition indicates that there is no instant dissipation of the mechanical energy in the thermal fluctuations [25, 27]. Because of the elastic component, the mechanical deformation energy is partly stored. As consequence, the internal energy of the liquid is not fully dissipated but increases upon applying a low frequency shear strain. We suspect that the same coupling process might occur applying a magnetic field. Indeed the complementary structural study carried out using neutron scattering in a wide scattering range does not have enabled to identify the emergence of a local order that nucleates and grows or a phase separation. In other words, correlations are not induced by an external field but might be preexistent which is in agreement with the identification of elastic correlations. In comparison with a previous study [24], increasing the alkyl chain length in ionic liquids seems to reinforce the shear elasticity which is also coherent with the observation of local layering structuration when the chain lengths increase [15-18].

How shear stress couples or competes with external fields (e.g. Maxwell magnetic stress) is challenging and calls in question the conventional hydrodynamic view. Recently several new theoretical developments have been independently emerged evidencing that liquids can support the propagation length of shear waves well above nanoscopic scales [27-30]. Therefore a new paradigm different from the low frequency viscous behavior imposed by the Maxwell model is emerging. The scale dependence of the shear elasticity is also predicted in particular in the frame of the non-affine models developed to quantitatively predict elastic and viscoelastic constants in glasses of polymers and colloids (NALD approach [31-35]). In another theoretical approach, authors substitute the description in terms of position and potential of molecular pair by a description in terms of wave vectors and energy densities with presents the advantage of representing the whole sample characteristics [36,37]. This approach takes into account various experimental results and successfully demonstrates the need to understand the laminar Newtonian regime as the asymptotic limit of the shear elasticity, in agreement with the transition observed in Fig. 1 and 2. A very recently statistical model rehabilitates the historical "forgetting" of the liquid shear waves and proposes a phononic approach in which elastic wave-packets interact with "liquid" particles [38]. This model integrates interestingly already the recent experimental observation of a liquid thermo-mechanical coupling [27].

About thirty years separate the pioneering experimental works of Derjaguin and co., interpreting the solid-like liquid behavior at the scale of several microns as an intrinsic liquid property [3] and mesoscopic measurements of the liquid elasticity [4-6, 24]. The existence of shear elasticity is also recognized by AFM [39] whose interpretation deviates from surface induced effects, by relaxation methods [40], by the identification of long range correlations via Rayleigh scattering [41] or by long range (bulk) electrostatic oscillations with the charge oscillations measured in ionic liquids [42], thus covering a very wide dimensional range and opening the way to new considerations about the nature of the small scale liquid state.

Acknowledgements:

The authors thank warmly Gregory Chaboussant and Marc Detrez for their help and advices about the neutron instrument PA20, and Denis Morineau, Bernard Doudin, Michael Coey for stimulating discussions about ionic liquids and paramagnetic liquids respectively. This work received funding from the European Union's Horizon 2020 research and innovation program under the Marie Sklodowska-Curie grant agreement no. 766007 and LabeX Palm (ANR-11-Idex-0003-02).